\documentclass[a4paper,aps,prl,superscriptaddress,nobibnotes,longbibliography,preprintnumbers,floatfix,twocolumn]{revtex4-2}
\usepackage[T1]{fontenc}
\usepackage[utf8]{inputenc}
\usepackage[english]{babel}
\usepackage[unicode, colorlinks=true, linkcolor=linkcolor, citecolor=linkcolor, filecolor=linkcolor, urlcolor=linkcolor, linktocpage, breaklinks]{hyperref}
\usepackage{csquotes,mathtools}
\usepackage[per-mode = symbol]{siunitx}
\usepackage{physics, tensor, bbm, mathrsfs}
\usepackage{amsmath, amssymb, amsfonts, amsthm, soul}
\usepackage{graphicx, comment, wasysym, placeins}
\usepackage[dvipsnames, usenames]{xcolor}
\usepackage{orcidlink}

\definecolor{linkcolor}{rgb}{0.0,0.3,0.5}
\definecolor{green}{rgb}{0.0,0.5,0.0}

\def\scri{\mathscr{I}}

\graphicspath{{./figures/}}

\newcommand{\textcomment}[1]{}

\begin{document}
\title{Exceptional Points and Resonance in Black Hole Ringdown}

\author{Rodrigo Panosso Macedo\orcidlink{0000-0003-2942-5080}}
    \affiliation{Center of Gravity, Niels Bohr Institute, Blegdamsvej 17, 2100 Copenhagen, Denmark}
\author{Takuya Katagiri\orcidlink{0000-0002-3755-3093}}
    \affiliation{Dipartimento di Fisica, “Sapienza” Universit`a di Roma \& Sezione INFN Roma1, P.A. Moro 5, 00185, Roma, Italy}
\author{Kei-ichiro Kubota\orcidlink{0000-0002-1576-4332}}
    \affiliation{Institute for Cosmic Ray Research, The University of Tokyo, 5-1-5 Kashiwanoha, Kashiwa, Chiba 277-8582, Japan}
\author{Hayato Motohashi\orcidlink{0000-0002-4330-7024}}
    \affiliation{Department of Physics, Tokyo Metropolitan University, 1-1 Minami-Osawa, Hachioji, Tokyo 192-0397, Japan}

\newcommand{\rpm}[1]{{\textcolor{blue}{\sf{[Rodrigo: #1]}}}}

\newcommand{\TK}[1]{{\textcolor{red}{\sf{[Takuya: #1]}}}}

\newcommand{\KK}[1]{{\textcolor{green}{\sf{[Kei-ichiro: #1]}}}}

\newcommand{\ha}[1]{{\textcolor{purple}{\sf{[Hayato: #1]}}}}

\begin{abstract}
We propose an exceptional-point (EP) framework for black-hole ringdown beyond the standard quasinormal-mode (QNM) paradigm. 
It provides a first-principles characterization of the resonance associated with avoided crossings near EPs, an effect that conventional QNM analysis cannot fully capture. 
Employing a phenomenological environmental black-hole model with the hyperboloidal framework, we identify near-coalescence of both QNM eigenvalues and eigenfunctions, and directly demonstrate that the resonance produces enhanced mode contributions in the time domain, resulting in characteristic departures from exponentially damped oscillations.
Our formulation further reveals that the EP frequency, given by the average of the resonant modes, emerges as the physically relevant observable in the near-EP regime, and offers a robust foundation for modeling and extracting resonant ringdown signals.
\end{abstract}

\maketitle

{\bf {\em Introduction.}}~Black holes (BHs) are macroscopic two-sided open systems: perturbations either fall irreversibly into the event horizon or radiate to infinity.
Such dissipative dynamics have long been central in physics, from Gamow's theory of $\alpha$ decay~\cite{Gamow:1928zz} and resonance scattering in particle physics~\cite{Taylor:1972pty,Kukulin1989} to open quantum systems~\cite{breuer2002theory}. 
More broadly, these phenomena fall under the framework of non-Hermitian physics, one of the major themes in condensed matter and photonics recently~\cite{moiseyev_2011,El-Ganainy2018,Bergholtz:2019deh,Ashida:2020dkc}. 
A hallmark of non-Hermitian systems is the existence of exceptional points (EPs)~\cite{Kato1995}, where not only eigenvalues but also their corresponding eigenvectors coalesce.
Mathematically, EPs appear as second-order poles of the Green's functions and underlie a variety of striking phenomena~\cite{Ashida:2020dkc,2019NatMa..18..783O,Wiersig:20,Parto2021,Ding:2022juv}, including characteristic departures from exponentially damped oscillations~\cite{PhysRevE.75.027201,Verstraete01032008,Heiss2010}.

Recent works have highlighted the non-Hermitian nature of BH ringdown, including pseudospectra and non-modal analysis~\cite{Jaramillo:2020tuu,Jaramillo:2022kuv}, resonance associated with avoided crossings~\cite{Motohashi:2024fwt}, and hysteresis~\cite{Cavalcante:2024swt,Cavalcante:2024kmy}.
In particular, the existence and implications of EPs in gravity were first identified in Ref.~\cite{Motohashi:2024fwt}, where the phenomenon appears between the fifth and sixth quadrupolar overtones of Kerr BHs.
Despite the observational challenges of accessing such higher overtones, avoided crossings can naturally arise in systems with sufficiently rich parameter spaces.
Indeed, similar mode-interaction phenomena occur in other astrophysical settings with multidimensional parameter spaces, such as relativistic stellar oscillations~\cite{1996ApJ...462..855A,Andersson:1997eq,Gondek:1997fd,1977A&A....58...41A,Sotani:2020eva}.

Notably, however, fifty years of BH ringdown studies have focused exclusively on quasinormal modes (QNMs) and excitation factors seen as first-order poles and residues of the Green's function, together with a more recent growing interest in the late-time tail decay and beyond-linear-regime interactions~\cite{Berti:2025hly}. 
This approach cannot capture double poles, even as interest in the near-EP resonance effect is rapidly growing~\cite{Motohashi:2024fwt,Cavalcante:2024kmy,Cavalcante:2024swt,Oshita:2025ibu,Lo:2025njp,Yang:2025dbn,Takahashi:2025uwo,Berti:2025hly,Chen:2025sbz,Kubota:2025hjk,Cavalcante:2025abr,Cao:2025afs}, motivated by the prospects of BH spectroscopy in the era of high-precision gravitational wave~(GW) astronomy~\cite{Berti:2025hly}. 

The lack of a systematic framework for BH spectroscopy including double-pole contributions, together with the universality of these phenomena, underscores the need to directly explore EPs and the non-Hermitian nature of gravity. To fill this gap, we propose a first-principles EP-based framework for BH ringdown that goes beyond the standard QNM paradigm. 
Central among the EP manifestations are resonances associated with avoided crossings, which together constitute a universal near-EP effect~\cite{Motohashi:2024fwt}. We characterize the distinct signatures of this effect and discuss their implications, including time-domain fitting analysis.
We employ geometrical units $c=G=1$.

{\bf {\em EP framework.}}~In linear perturbation theory, the ringdown waveform of perturbed BHs observed at large distances can be expressed as a frequency-domain integral of the Green's function against initial data~\cite{Leaver:1986gd}, schematically captured by a function $A(\omega)$ as
\begin{align}\label{eq:integral_Psi}
\Psi = \int_{-\infty}^\infty \frac{d\omega}{2\pi i} \frac{e^{-i\omega t}}{A(\omega)}.
\end{align}
In the conventional treatment, the poles of the integrand are assumed to be simple: 
$A(\omega)\approx (\omega-\omega_n) A'(\omega_n)$ 
with $\omega_n$ the complex QNM frequencies. 
Closing the contour in the complex plane yields the usual QNM superposition,
\begin{align}\label{eq:QNM_Psi}
\Psi = -\sum_n \frac{e^{-i\omega_n t}}{A'(\omega_n)}.
\end{align}
where the coefficients $1/A'(\omega_n)$ correspond to excitation factors multiplied by initial-data-dependent terms.
Here, we neglect power-law tails for simplicity.

This simple-pole analysis, however, breaks down in the presence of double poles, i.e., EPs. 
Near an EP, 
$A(\omega)\approx (\omega-\omega_\mathrm{EP})^2 A''(\omega_\mathrm{EP})/2$, 
leading to a qualitatively distinct time dependence,
\begin{align}
\Psi &= \left( \frac{A'''(\omega_\mathrm{EP})}{ 3 A''(\omega_\mathrm{EP})} +it \right) \frac{2e^{-i\omega_\mathrm{EP} t}}{A''(\omega_\mathrm{EP})}
-\sum_{n\ne \mathrm{EP}} \frac{e^{-i\omega_n t}}{A'(\omega_n)} .
\label{eq:EP_Psi}
\end{align}
In contrast to the conventional QNM framework~\eqref{eq:QNM_Psi}, the EP contribution carries a polynomial prefactor in time.
While systems typically exhibit sharp avoided crossings rather than exact degeneracies in practice~\cite{Onozawa:1996ux,Yang:2013uba,Cook:2014cta,Jansen:2017oag,Dias:2021yju,Davey:2022vyx,Dias:2022oqm,Kinoshita:2023iad,Motohashi:2024fwt,Cavalcante:2024kmy,Cavalcante:2024swt,Oshita:2025ibu,Lo:2025njp,Yang:2025dbn,Takahashi:2025uwo,Berti:2025hly,Chen:2025sbz,Kubota:2025hjk,Cavalcante:2025abr,Cao:2025afs},
the EP formalism exposes the underlying essential structure, providing an excellent approximation in their vicinity.

To make contact with the conventional language of QNM excitation factors, we next focus on near-EP dynamics dominated by an avoided crossing and its associated resonance between two overtones $j$ and $k$ as a system parameter $p$ varies.
This effect manifests as hyperbolic trajectories for $\omega$ and lemniscate trajectories for the excitation factors $B$ in the complex plane~\cite{Motohashi:2024fwt}, with their averages
$\omega_c \coloneqq (\omega_j+\omega_k)/2$ and $B_c \coloneqq (B_j+B_k)/2$ 
remaining nearly constant with respect to $p$. 
For a sharp avoided crossing, $\omega_c\approx \omega_\mathrm{EP}$ holds.
We further denote $B_\mathrm{EP}=2B_c$.
The frequencies and excitation factors can then be parameterized as
\begin{align}
\omega_{j,k} = \omega_\mathrm{EP} [1\pm \delta(p)], \quad
B_{j,k} = B_\mathrm{EP} \left[ 1 \pm b/\delta(p) \right] /2,
\end{align}
where $b$ is a complex constant and $\delta(p)$ traces a hyperbola in the complex plane, remaining small for a sharp avoided crossing.

With this parameterization, for $|t \omega_\mathrm{EP} \delta|\ll 1$, the resonant contribution to the time-domain signal reduces to
\begin{align}
B_j e^{-i\omega_j t} + B_k e^{-i\omega_k t} \simeq B_\mathrm{EP} ( 1 - i b \omega_\mathrm{EP} t ) e^{-i\omega_\mathrm{EP} t} .
\label{eq:fittingfunc}
\end{align}
This directly links the EP contribution in Eq.~\eqref{eq:EP_Psi} to the conventional QNM description.
While Ref.~\cite{Yang:2025dbn} suggests a similar term in $te^{-i\omega_j t}$, Eq.~\eqref{eq:fittingfunc} elucidates the underlying physics by:
(i) rigorously tracing the linear term to a second-order pole; 
(ii) identifying the EP frequency $\omega_\mathrm{EP}$, arising from the averaged resonant QNMs, as the central observable in waveform modeling; and 
(iii) retaining from first principles the additional constant-amplitude damped sinusoid with the EP frequency $\omega_\mathrm{EP}$.

We apply the above framework to a phenomenological model of a BH surrounded by a localized matter environment, which serves as a setting to probe various aspects of EPs in BH spectroscopy.

\begin{figure*}[ht!]
\centering
\includegraphics[width=0.88\textwidth]{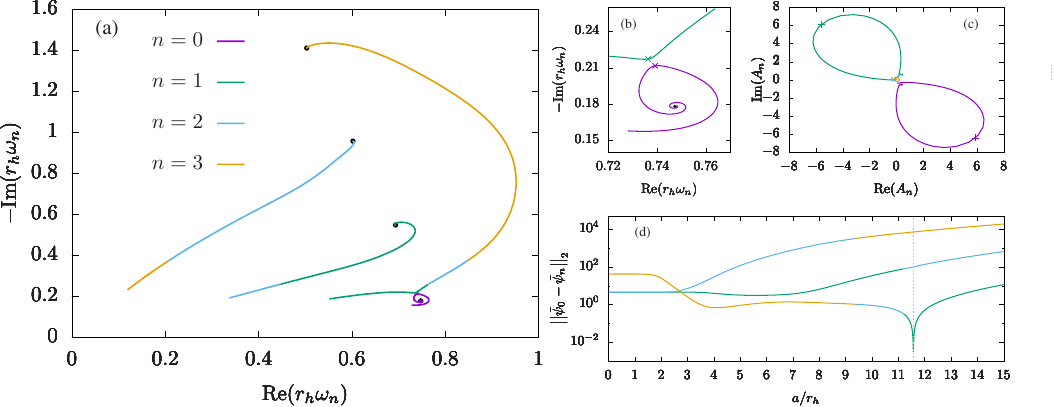}
\caption{{\em Panels (a) and (b):} Migration of the QNMs for parameter range $a/r_h\in[0,15]$ with $\epsilon=\epsilon_*(\simeq 0.00204 \,r_h^{-2})$. 
The fundamental mode $n=0$ (purple) and the overtones $n=1$--$3$ (green, blue, yellow) are displayed, with black markers denoting the Schwarzschild values ($\epsilon=0$). 
Overtaken transitions occur between overtones and the avoided crossing between the resonant $n=0$ and $n=1$ (originally $n=3$) modes; see the zoom in Panel (b) around $a=a_*(\simeq 11.5722\, r_h)$ (cross markers). 
{\em Panel (c):} Amplitude trajectories of the individual tones of the time-domain signals, reconstructed from constant initial data and measured at future null infinity. 
The fundamental mode and the first overtone trace lemniscate (figure-eight-like) trajectories, attaining maxima at $a=a_*$ (plus markers). 
{\em Panel (d):} $L^2$-norm $\left\Vert \bar \psi_0(\sigma) - \bar \psi_n(\sigma) \right\Vert_2$ between the fundamental and overtone QNM eigenfunctions. 
Eigenfunctions $n=0$ and $n=1$ nearly coincide at $a=a_*$ (dashed line), confirming an EP signature.}
\label{fig:QNM_AV}
\end{figure*}

{\bf{\em Environmental BH model.}}~BH perturbation theory typically casts gravitational perturbations into a set of second-order differential wave equations for gauge-invariant variables~\cite{Chandrasekhar:1985kt}. 
The QNM damped oscillations arise as an infinite set of complex eigenvalues $\omega_n$ associated with eigenfunctions satisfying no outgoing waves at the horizon and no incoming waves at spatial infinity.

For our purpose, it suffices to focus on spherically symmetric spacetimes. After projection into angular harmonics~\footnote{We omit the angular indices $(\ell,m)$ for simplicity.} and employing Schwarzschild coordinates $x^\mu=(t,r)$, the GW signal is described by a master function $\Psi(t,r)$, or its frequency-domain projection $\Psi(t,r)\sim e^{-i\omega t}\psi(r)$, satisfying
\begin{align}
\label{eq:RWeq}
    -\dfrac{d^2 \Psi}{dt^2}+\frac{d^2 \Psi}{dr_*^2}-V\Psi=0, \quad 
    \frac{d^2 \psi}{dr_*^2}+\left(\omega^2-V\right)\psi=0.
\end{align}
The tortoise coordinate $r_*$ is defined by $dr_*/dr=1/f(r)$ with $f(r)=1-{r_h}/{r}$, where $r_h$ is the horizon radius.

Although exceptional points are expected to arise naturally in systems with richer parameter spaces, it is noteworthy that setups capturing environmental effects in a minimal setting already suffice to produce resonant dynamics~\cite{Motohashi:2024fwt}. We consider a potential $V(r)$ of the form
\begin{equation}
    \label{eq:Potential}
    V(r) = f(r)\!\left(\frac{\ell(\ell+1)}{r^2}-\frac{3r_h}{r^3}\right) + \dfrac{\epsilon}{ \cosh^{2}\!\left(\frac{r_* -a}{r_h} \right)}.
\end{equation}
The first term is the standard axial perturbation potential.
We set $\ell=2$, corresponding to the dominant quadrupole contribution to the GW signal.  
The second term is a modification adding a small ``bump'' with amplitude $\epsilon$, modeling a localized matter shell at $r_*=a$~\cite{Cheung:2021bol}. 

To solve Eq.~\eqref{eq:RWeq} by geometrically imposing outgoing boundary conditions, we employ the hyperboloidal framework~\cite{Zenginoglu:2011jz,PanossoMacedo:2023qzp,PanossoMacedo:2024nkw}.
A new coordinate system $\bar x^\mu=(\tau,\sigma)$ compactifies the radial direction via $r={r_h}/{\sigma}$. 
The compactification is performed along new time hypersurfaces $\tau={\rm const.}$ generated by deforming the original time coordinate $t/r_h$ with a height function $H(\sigma) = -{1}/{\sigma} + \ln\left(\sigma(1-\sigma) \right)$. 
Within this framework, $\sigma=0,1$ correspond to future null infinity and the BH horizon, respectively. 
The time- and frequency-domain master functions are mapped into $\bar\Psi(\tau,\sigma) = \Psi(t(\tau,\sigma),r(\sigma))$ and $\bar \psi(\sigma) = e^{i r_h \omega H(\sigma)}\,\psi(r(\sigma))$, with further details of the resulting equation given in Ref.~\cite{PanossoMacedo:2023qzp}. 

A unique solution to the hyperboloidal wave equation requires only prescription of initial data. We adopt static initial data with a constant field on the initial slice,
$(\bar\Psi(0,\sigma),\partial_\tau\bar\Psi(0,\sigma))=(1,0)$.
This simple setup yields GW content uniformly distributed from the BH horizon to future null infinity $\scri^+$, directly accessing the QNM ringdown dynamics without the prompt response~\cite{Ansorg:2016ztf}.

The time- and frequency-domain equations~\eqref{eq:RWeq} are then solved with a multi-domain Chebyshev spectral method~\cite{PanossoMacedo:2014dnr,Jaramillo:2020tuu,Bourg:2025lpd}, with numerical accuracy enhanced with analytical mesh-refinement techniques~\cite{PanossoMacedo:2022fdi,Zhou:2025xta}. 
By combining frequency-domain amplitudes obtained directly from QNM eigenfunctions~\cite{Ansorg:2016ztf,Bourg:2025lpd} with accurate time-domain fits, this approach offers a robust methodology for the systematic exploration of EPs.

{\bf {\em QNM coalescence and resonance.}}~The model~\eqref{eq:Potential} is known to destabilize the QNM spectrum~\cite{Cheung:2021bol} up to the fundamental mode, even for very small bump amplitudes $\epsilon\, r_h^2 \ll 1$, while leaving the early-time ringdown largely unaffected~\cite{Berti:2022xfj,Kyutoku:2022gbr}. 
Most importantly, the model naturally yields resonant excitations at avoided crossings~\cite{Motohashi:2024fwt}, as clearly portrayed in Fig.~\ref{fig:QNM_AV}.

Figure~\ref{fig:QNM_AV}(a) shows the trajectory of the fundamental mode $n=0$ and overtones $n=1$--$3$, ordered by $|{\rm Im}(r_h \omega_n)|$. 
As $a$ increases, higher overtones can ``overtake'' lower ones, while lower overtones transition into higher ones. 
Besides, modes can approach each other up to a critical value $a=a_*$, and then repel after passing this point. 
This anomalous behavior at $a=a_*$ is an avoided crossing.

For instance, while the initially fundamental mode retains the lowest decay rate, the initially third overtone transitions into the second overtone at $a \simeq 9.40\, r_h$ and into the first overtone at $a \simeq 11.21\, r_h$. 
At $a=a_*(\simeq 11.5722\, r_h)$, the $n=0$ and $n=1$ QNMs undergo an avoided crossing, as highlighted in Fig.~\ref{fig:QNM_AV}(b).

Figure~\ref{fig:QNM_AV}(c) displays the strong amplification of the fundamental mode and first overtone amplitudes around $a=a_*$. Remarkably, their trajectories closely follow the lemniscate pattern predicted by the excitation factors~\cite{Motohashi:2024fwt}, providing the first direct theoretical demonstration that the resonance encoded in the excitation factors also manifests in the QNM amplitudes generated by nontrivial initial data. 

In the near-EP regime, we expect that the qualitative behavior of the QNM amplitudes is the same as that of the excitation factors, regardless of the choice of initial data. 
The QNM amplitudes depend on the excitation factors, the initial data, and the QNM eigenfunctions. 
When the eigenfunctions nearly coincide near an EP, the excitation-factor contribution dominates the QNM amplitudes, ensuring the persistence of the resonance.

To verify the eigenfunctions' behavior, we leverage their regularity within the hyperboloidal framework. Fig.~\ref{fig:QNM_AV}(d) presents the $L^2$-norm $\left\Vert \bar \psi_0(\sigma) - \bar \psi_n(\sigma) \right\Vert_2$ between each overtone and the fundamental mode.
A key result is that the $L^2$-norm between the $n=0$ and $n=1$ modes drops significantly at $a=a_*$, indicating that the two eigenfunctions do indeed become nearly identical.
This is an explicit manifestation of {\it near-coalescence of eigenfunctions} at EPs~\cite{Kato1995}, thereby confirming the non-Hermitian nature of BH perturbation theory~\footnote{While finishing this work, Ref.~\cite{Cao:2025afs} also comments on a similar result.}.

\begin{figure}[ht!]
\centering
\includegraphics[width=0.42\textwidth]{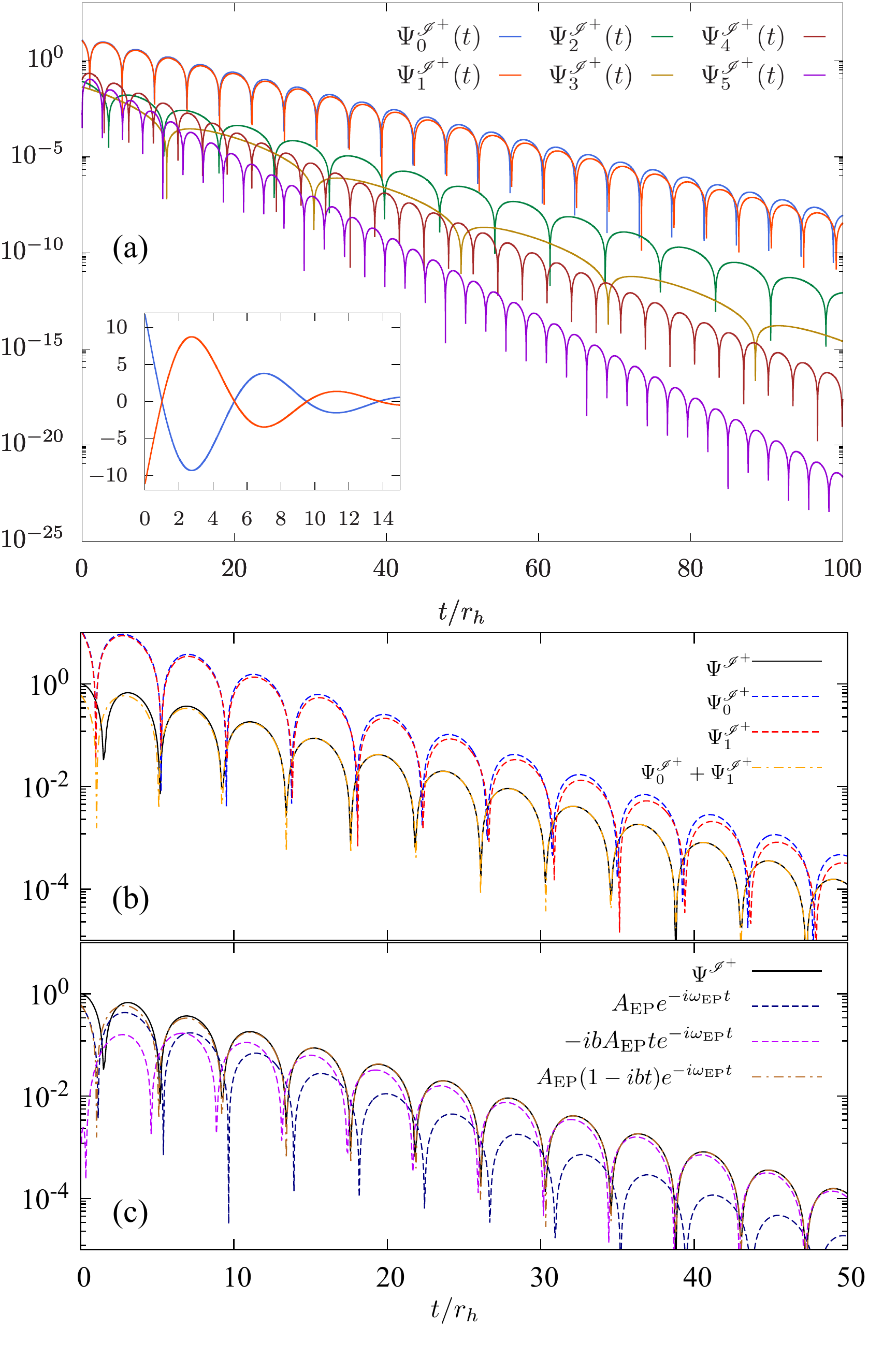}
\caption{Wave signal at future null infinity $\scri^+$ for $(\epsilon, a)=(\epsilon_*, a_*)$. 
{\em Panel (a):} Decomposition of the reconstructed resonant waveform that evolves from the constant initial data~$(\Psi,\partial_\tau\Psi)|_{\tau=0}=(1,0)$.
The fundamental mode and first overtone are significantly excited compared to the others. 
The inset highlights the fundamental mode and first overtone, showing they are nearly out of phase. 
{\em Panel (b):} Comparison among the full signal~(solid black), the reconstructed fundamental mode~(dashed blue) and first overtone~(dashed red), and their superposition~(dash-dotted orange). 
{\em Panel (c):} Comparison among the full signal~(solid black), the reconstructed constant amplitude term~(dashed navy) and linearly growing term~(dashed dark magenta) in the EP model~\eqref{eq:fittingfunc} and their superposition~(dash-dotted dark red).
}
\label{fig:TimeEvol}
\end{figure}

{\bf {\em Resonant ringdown waveform.}} 
We next turn to explore the time-domain waveform $\Psi^{\scri^+}(t)$ observed at future null infinity $\scri^+$, which reveals how the resonant signature appears in the signal.
Figure~\ref{fig:TimeEvol}(a) shows the contributions from each overtone, reconstructed using the frequency-domain techniques~\cite{Ansorg:2016ztf,Bourg:2025lpd}. The fundamental mode and first overtone are significantly larger than the higher overtones. 
Consistent with the lemniscate complex amplitude trajectories in Fig.~\ref{fig:QNM_AV}(c), the inset confirms they have almost opposite phases. 

Based on the QNM paradigm, such resonant destructive interference should not cause obvious signatures in the ringdown signal~\cite{Oshita:2025ibu}. 
Figure~\ref{fig:TimeEvol}(b) compares the full signal with the reconstructed $n=0$ and $n=1$ contributions.
While individually they exceed the full signal in amplitude, their sum reproduces its dominant behavior.

We now uncover the underlying structure of the signal through the EP formalism.
Figure~\ref{fig:TimeEvol}(c) compares the contributions from each term in Eq.~\eqref{eq:fittingfunc} individually.
Their evolution is obtained from amplitudes and frequencies calculated directly from the reconstructed $n=0$ and $n=1$ components, i.e., without any fitting, arising directly from the semi-analytical frequency-domain analysis. 
As expected, the damped-oscillatory term alone fails to capture the dynamics, whereas the linear term individually reproduces the waveform
only at later times $t\gtrsim 25\,r_h$. 
However, as predicted by the EP analysis \eqref{eq:fittingfunc}, their sum captures the resonant waveform accurately across all times.

This semi-analytical comparison already shows that the first-principles EP formula~\eqref{eq:fittingfunc} robustly captures the resonant waveform in the near-EP regime, motivating the fitting analysis that follows.

{\bf {\em Fitting.}}~Complementing the above semi-analytical results, we perform a time-domain fitting of the waveform using \texttt{NonlinearModelFit} in \texttt{Mathematica} with the following models:
(i) \texttt{2QNM}: simple-pole analysis $\Psi^{\scri^+}(t)=\sum_{i=0}^1 C_i e^{-i\omega_i t}$; 
(ii) \texttt{EP Linear}: EP linear growth $\Psi^{\scri^+}(t)=D_{\rm EP}\, t\, e^{-i\omega_{\rm EP} t}$; and 
(iii) \texttt{EP total}: complete EP model $\Psi^{\scri^+}(t)=(C_{\rm EP} + D_{\rm EP} t) e^{-i\omega_{\rm EP}\, t}$. 
After removing the late-time tail via an asymptotic power-law fit~\cite{Kubota:2025hjk}, we identify reliable ringdown dynamics up to $t/r_h\simeq 130$. 
Varying the fitting start time within $t_i/r_h\in[0,100]$, we extract QNM frequencies at minimal mismatch, typically for $t_i/r_h\sim 25$--$40$. 

\begin{figure}[t!]
    \centering
    \includegraphics[width=0.48\textwidth]{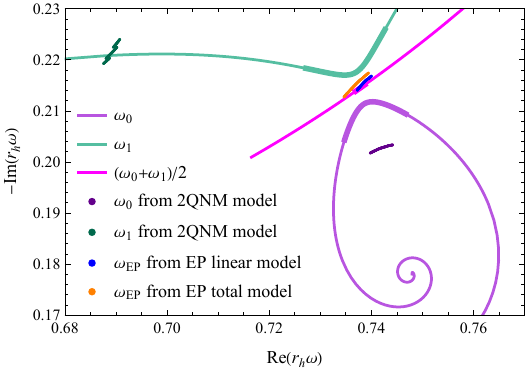}
    \caption{
    Extracted frequencies obtained using the \texttt{2QNM}~(purple and green dots), \texttt{EP linear}~(blue dots), and \texttt{EP total} models~(orange dots) for bump positions $a/r_h\in[11.54,11.6]$ with $\epsilon = \epsilon_*$. 
    The curves denote the corresponding QNM trajectories for the fundamental mode~(light purple), first overtone~(light green), and their average~(magenta), with thicker segments indicating the parameter range where the fitting is performed.}
    \label{fig:fitting}
\end{figure}

Figure~\ref{fig:fitting} expands on Fig.~\ref{fig:QNM_AV}(b) by including the average $\omega_c=(\omega_0+\omega_1)/2$.
The thicker segments of the purple, green, and magenta curves highlight the parameter range $a/r_h\in[11.54,11.6]$ where the fitting is performed. 
Confirming the theoretical prediction~\cite{Motohashi:2024fwt}, $\omega_c$ stays nearly constant in this hyperbolic regime, lying close to the EP frequency.
Figure~\ref{fig:fitting} also shows the extracted frequencies according to the three models. 
The \texttt{2QNM} model fails completely to recover the correct $n=0$ and $n=1$ QNM frequencies, whereas the EP models correctly fit $\omega_{\rm c}$ in accordance with the EP framework.
In particular, the \texttt{EP Linear} model shows better overlap with the analytical EP frequency. 
The \texttt{EP total} model shows a slightly larger spread but achieves a notable improvement in mismatch by a factor of order unity, likely due to its additional fitting parameters. 

Moreover, the \texttt{2QNM} model is unstable within our fitting strategy. 
In Fig.~\ref{fig:fitting}, where the initial seeds are set to $\omega_0$ and $\omega_1$, the \texttt{2QNM} fit already drifts far from the correct QNM values. 
This behavior worsens when the initial seeds are replaced by the original $n=0,1$ Schwarzschild frequencies.
By contrast, the EP models remain robust: although Fig.~\ref{fig:fitting} uses $\omega_c$ as initial seeds, we also recover the expected EP frequency when starting from the Schwarzschild fundamental mode.
These results confirm the EP signatures and highlight the advantage of first-principles EP models in fitting strategies.

{\bf {\em Conclusion and discussion.}}~Extending the conventional QNM paradigm, we propose an EP framework for BH ringdown based on first-principles calculations that incorporate the non-Hermitian structure of the problem. Motivated by the recent surge of interest in EP-related phenomena in gravitational physics~\cite{Motohashi:2024fwt,Cavalcante:2024kmy,Cavalcante:2024swt,Oshita:2025ibu,Lo:2025njp,Yang:2025dbn,Takahashi:2025uwo,Berti:2025hly,Chen:2025sbz,Kubota:2025hjk,Cavalcante:2025abr,Cao:2025afs}, we investigate the resonance associated with avoided crossings near EPs~\cite{Motohashi:2024fwt}, analyzing its frequency- and time-domain signatures using a phenomenological model of a BH interacting with environmental matter.

Our analysis demonstrates the resonant amplification of QNM contributions in the time domain and the near-coalescence of eigenfunctions, providing a clear manifestation of non-Hermitian physics in gravity. The hyperboloidal framework plays a central role by casting the QNM problem into a well-defined Hilbert space~\cite{Zenginoglu:2011jz,PanossoMacedo:2024nkw,Gajic:2024xrn}, where eigenfunctions remain regular and finite in the exterior region, allowing the characteristic eigenfunction coalescence of EPs to be demonstrated explicitly. We further show that the EP framework accurately captures the resonant waveform over the relevant timescales. In particular, the EP frequency—given by the average of the two resonant QNM frequencies in the near-EP regime—emerges as the physically meaningful observable, enabling substantially more robust waveform fitting than conventional QNM models.

Although avoided crossings may appear to require fine tuning, they arise naturally in systems with sufficiently rich parameter spaces, as is known, for example, from relativistic stellar oscillations~\cite{1996ApJ...462..855A,Andersson:1997eq,Gondek:1997fd,1977A&A....58...41A,Sotani:2020eva}. Remarkably, the phenomenon already emerges in the simple setup considered here, where environmental effects are modeled through a minimal modification of the vacuum BH potential. This suggests that EP signatures should be expected in realistic astrophysical scenarios where environmental effects or beyond-GR physics enlarge the parameter space accessible to GW observations.

These results highlight the importance of waveform models derived from first principles that incorporate the non-Hermitian structure of the problem. 
This perspective places EP dynamics at the foundation of resonant BH ringdown. 
Such modeling will be essential for developing robust data-analysis strategies capable of capturing the characteristic time-domain behavior associated with EP dynamics in future high-precision GW observations.

{\bf {\em Acknowledgments.}} 
R.P.M.\ and T.K.\ acknowledge support from the Villum Investigator program supported by the VILLUM Foundation (grant no.\ VIL37766) and the DNRF Chair program (grant no.\ DNRF162) by the Danish National Research Foundation. 
This work has received funding from the European Union's Horizon 2020 research and innovation programme under the Marie Sklodowska-Curie grant agreement No.\ 101131233. The Center of Gravity is a Center of Excellence funded by the Danish National Research Foundation under grant No.\ 184.
T.K.\ is supported by the MUR FIS2 Advanced Grant ET-NOW (CUP:~B53C25001080001) and by the INFN TEONGRAV initiative. 
H.M.\ was supported in part by Japan Society for the Promotion of Science (JSPS) Grant-in-Aid for Scientific Research (KAKENHI) Grant No.~JP22K03639. Codes available in \cite{rodrigo_panosso_macedo_2026_20177185}.

\bibliography{Ref}

\begin{thebibliography}{60}%
\makeatletter
\providecommand \@ifxundefined [1]{%
 \@ifx{#1\undefined}
}%
\providecommand \@ifnum [1]{%
 \ifnum #1\expandafter \@firstoftwo
 \else \expandafter \@secondoftwo
 \fi
}%
\providecommand \@ifx [1]{%
 \ifx #1\expandafter \@firstoftwo
 \else \expandafter \@secondoftwo
 \fi
}%
\providecommand \natexlab [1]{#1}%
\providecommand \enquote  [1]{``#1''}%
\providecommand \bibnamefont  [1]{#1}%
\providecommand \bibfnamefont [1]{#1}%
\providecommand \citenamefont [1]{#1}%
\providecommand \href@noop [0]{\@secondoftwo}%
\providecommand \href [0]{\begingroup \@sanitize@url \@href}%
\providecommand \@href[1]{\@@startlink{#1}\@@href}%
\providecommand \@@href[1]{\endgroup#1\@@endlink}%
\providecommand \@sanitize@url [0]{\catcode `\\12\catcode `\$12\catcode
  `\&12\catcode `\#12\catcode `\^12\catcode `\_12\catcode `\%12\relax}%
\providecommand \@@startlink[1]{}%
\providecommand \@@endlink[0]{}%
\providecommand \url  [0]{\begingroup\@sanitize@url \@url }%
\providecommand \@url [1]{\endgroup\@href {#1}{\urlprefix }}%
\providecommand \urlprefix  [0]{URL }%
\providecommand \Eprint [0]{\href }%
\providecommand \doibase [0]{http://dx.doi.org/}%
\providecommand \selectlanguage [0]{\@gobble}%
\providecommand \bibinfo  [0]{\@secondoftwo}%
\providecommand \bibfield  [0]{\@secondoftwo}%
\providecommand \translation [1]{[#1]}%
\providecommand \BibitemOpen [0]{}%
\providecommand \bibitemStop [0]{}%
\providecommand \bibitemNoStop [0]{.\EOS\space}%
\providecommand \EOS [0]{\spacefactor3000\relax}%
\providecommand \BibitemShut  [1]{\csname bibitem#1\endcsname}%
\let\auto@bib@innerbib\@empty
\bibitem [{\citenamefont {Gamow}(1928)}]{Gamow:1928zz}%
  \BibitemOpen
  \bibfield  {author} {\bibinfo {author} {\bibfnamefont {G.}~\bibnamefont
  {Gamow}},\ }\bibfield  {title} {\enquote {\bibinfo {title} {{Zur
  Quantentheorie des Atomkernes}},}\ }\href {\doibase 10.1007/BF01343196}
  {\bibfield  {journal} {\bibinfo  {journal} {Z. Phys.}\ }\textbf {\bibinfo
  {volume} {51}},\ \bibinfo {pages} {204} (\bibinfo {year} {1928})}\BibitemShut
  {NoStop}%
\bibitem [{\citenamefont {Taylor}(1972)}]{Taylor:1972pty}%
  \BibitemOpen
  \bibfield  {author} {\bibinfo {author} {\bibfnamefont {J.~R.}\ \bibnamefont
  {Taylor}},\ }\href@noop {} {\emph {\bibinfo {title} {{Scattering Theory: The
  Quantum Theory of Nonrelativistic Collisions}}}}\ (\bibinfo  {publisher}
  {John Wiley {\&} Sons, Inc.},\ \bibinfo {address} {New York},\ \bibinfo
  {year} {1972})\BibitemShut {NoStop}%
\bibitem [{\citenamefont {Kukulin}\ \emph {et~al.}(1989)\citenamefont
  {Kukulin}, \citenamefont {Krasnopol'sky},\ and\ \citenamefont
  {Hor\'{a}\v{c}ek}}]{Kukulin1989}%
  \BibitemOpen
  \bibfield  {author} {\bibinfo {author} {\bibfnamefont {V.~I.}\ \bibnamefont
  {Kukulin}}, \bibinfo {author} {\bibfnamefont {V.~M.}\ \bibnamefont
  {Krasnopol'sky}}, \ and\ \bibinfo {author} {\bibfnamefont {J.}~\bibnamefont
  {Hor\'{a}\v{c}ek}},\ }\href {\doibase 10.1007/978-94-015-7817-2} {\emph
  {\bibinfo {title} {Theory of Resonances: Principles and Applications}}},\
  Reidel Texts in the Mathematical Sciences\ (\bibinfo  {publisher} {Springer
  Dordrecht},\ \bibinfo {year} {1989})\BibitemShut {NoStop}%
\bibitem [{\citenamefont {Breuer}\ and\ \citenamefont
  {Petruccione}(2002)}]{breuer2002theory}%
  \BibitemOpen
  \bibfield  {author} {\bibinfo {author} {\bibfnamefont {H.}~\bibnamefont
  {Breuer}}\ and\ \bibinfo {author} {\bibfnamefont {F.}~\bibnamefont
  {Petruccione}},\ }\href {https://books.google.co.jp/books?id=0Yx5VzaMYm8C}
  {\emph {\bibinfo {title} {The Theory of Open Quantum Systems}}}\ (\bibinfo
  {publisher} {Oxford University Press},\ \bibinfo {year} {2002})\BibitemShut
  {NoStop}%
\bibitem [{\citenamefont {Moiseyev}(2011)}]{moiseyev_2011}%
  \BibitemOpen
  \bibfield  {author} {\bibinfo {author} {\bibfnamefont {N.}~\bibnamefont
  {Moiseyev}},\ }\href {\doibase 10.1017/CBO9780511976186} {\emph {\bibinfo
  {title} {Non-Hermitian Quantum Mechanics}}}\ (\bibinfo  {publisher}
  {Cambridge University Press},\ \bibinfo {year} {2011})\BibitemShut {NoStop}%
\bibitem [{\citenamefont {El-Ganainy}\ \emph {et~al.}(2018)\citenamefont
  {El-Ganainy}, \citenamefont {Makris}, \citenamefont {Khajavikhan},
  \citenamefont {Musslimani}, \citenamefont {Rotter},\ and\ \citenamefont
  {Christodoulides}}]{El-Ganainy2018}%
  \BibitemOpen
  \bibfield  {author} {\bibinfo {author} {\bibfnamefont {R.}~\bibnamefont
  {El-Ganainy}}, \bibinfo {author} {\bibfnamefont {K.~G.}\ \bibnamefont
  {Makris}}, \bibinfo {author} {\bibfnamefont {M.}~\bibnamefont {Khajavikhan}},
  \bibinfo {author} {\bibfnamefont {Z.~H.}\ \bibnamefont {Musslimani}},
  \bibinfo {author} {\bibfnamefont {S.}~\bibnamefont {Rotter}}, \ and\ \bibinfo
  {author} {\bibfnamefont {D.~N.}\ \bibnamefont {Christodoulides}},\ }\bibfield
   {title} {\enquote {\bibinfo {title} {{Non-Hermitian physics and PT
  symmetry}},}\ }\href {\doibase 10.1038/nphys4323} {\bibfield  {journal}
  {\bibinfo  {journal} {Nature Physics}\ }\textbf {\bibinfo {volume} {14}},\
  \bibinfo {pages} {11} (\bibinfo {year} {2018})}\BibitemShut {NoStop}%
\bibitem [{\citenamefont {Bergholtz}\ \emph {et~al.}(2021)\citenamefont
  {Bergholtz}, \citenamefont {Budich},\ and\ \citenamefont
  {Kunst}}]{Bergholtz:2019deh}%
  \BibitemOpen
  \bibfield  {author} {\bibinfo {author} {\bibfnamefont {E.~J.}\ \bibnamefont
  {Bergholtz}}, \bibinfo {author} {\bibfnamefont {J.~C.}\ \bibnamefont
  {Budich}}, \ and\ \bibinfo {author} {\bibfnamefont {F.~K.}\ \bibnamefont
  {Kunst}},\ }\bibfield  {title} {\enquote {\bibinfo {title} {{Exceptional
  topology of non-Hermitian systems}},}\ }\href {\doibase
  10.1103/revmodphys.93.015005} {\bibfield  {journal} {\bibinfo  {journal}
  {Rev. Mod. Phys.}\ }\textbf {\bibinfo {volume} {93}},\ \bibinfo {pages}
  {015005} (\bibinfo {year} {2021})},\ \Eprint
  {http://arxiv.org/abs/1912.10048} {arXiv:1912.10048 [cond-mat.mes-hall]}
  \BibitemShut {NoStop}%
\bibitem [{\citenamefont {Ashida}\ \emph {et~al.}(2021)\citenamefont {Ashida},
  \citenamefont {Gong},\ and\ \citenamefont {Ueda}}]{Ashida:2020dkc}%
  \BibitemOpen
  \bibfield  {author} {\bibinfo {author} {\bibfnamefont {Y.}~\bibnamefont
  {Ashida}}, \bibinfo {author} {\bibfnamefont {Z.}~\bibnamefont {Gong}}, \ and\
  \bibinfo {author} {\bibfnamefont {M.}~\bibnamefont {Ueda}},\ }\bibfield
  {title} {\enquote {\bibinfo {title} {{Non-Hermitian physics}},}\ }\href
  {\doibase 10.1080/00018732.2021.1876991} {\bibfield  {journal} {\bibinfo
  {journal} {Adv. Phys.}\ }\textbf {\bibinfo {volume} {69}},\ \bibinfo {pages}
  {249} (\bibinfo {year} {2021})},\ \Eprint {http://arxiv.org/abs/2006.01837}
  {arXiv:2006.01837 [cond-mat.mes-hall]} \BibitemShut {NoStop}%
\bibitem [{\citenamefont {Kato}(1995)}]{Kato1995}%
  \BibitemOpen
  \bibfield  {author} {\bibinfo {author} {\bibfnamefont {T.}~\bibnamefont
  {Kato}},\ }\href {\doibase 10.1007/978-3-642-66282-9} {\emph {\bibinfo
  {title} {Perturbation Theory for Linear Operators}}}\ (\bibinfo  {publisher}
  {Springer-Verlag, Berlin},\ \bibinfo {year} {1995})\BibitemShut {NoStop}%
\bibitem [{\citenamefont {{{\"O}zdemir}}\ \emph {et~al.}(2019)\citenamefont
  {{{\"O}zdemir}}, \citenamefont {{Rotter}}, \citenamefont {{Nori}},\ and\
  \citenamefont {{Yang}}}]{2019NatMa..18..783O}%
  \BibitemOpen
  \bibfield  {author} {\bibinfo {author} {\bibfnamefont {{\c{S}}.~K.}\
  \bibnamefont {{{\"O}zdemir}}}, \bibinfo {author} {\bibfnamefont
  {S.}~\bibnamefont {{Rotter}}}, \bibinfo {author} {\bibfnamefont
  {F.}~\bibnamefont {{Nori}}}, \ and\ \bibinfo {author} {\bibfnamefont
  {L.}~\bibnamefont {{Yang}}},\ }\bibfield  {title} {\enquote {\bibinfo {title}
  {{Parity-time symmetry and exceptional points in photonics}},}\ }\href
  {\doibase 10.1038/s41563-019-0304-9} {\bibfield  {journal} {\bibinfo
  {journal} {Nature Materials}\ }\textbf {\bibinfo {volume} {18}},\ \bibinfo
  {pages} {783} (\bibinfo {year} {2019})}\BibitemShut {NoStop}%
\bibitem [{\citenamefont {Wiersig}(2020)}]{Wiersig:20}%
  \BibitemOpen
  \bibfield  {author} {\bibinfo {author} {\bibfnamefont {J.}~\bibnamefont
  {Wiersig}},\ }\bibfield  {title} {\enquote {\bibinfo {title} {Review of
  exceptional point-based sensors},}\ }\href {\doibase 10.1364/PRJ.396115}
  {\bibfield  {journal} {\bibinfo  {journal} {Photon. Res.}\ }\textbf {\bibinfo
  {volume} {8}},\ \bibinfo {pages} {1457} (\bibinfo {year} {2020})}\BibitemShut
  {NoStop}%
\bibitem [{\citenamefont {Parto}\ \emph {et~al.}(2021)\citenamefont {Parto},
  \citenamefont {Liu}, \citenamefont {Bahari}, \citenamefont {Khajavikhan},\
  and\ \citenamefont {Christodoulides}}]{Parto2021}%
  \BibitemOpen
  \bibfield  {author} {\bibinfo {author} {\bibfnamefont {M.}~\bibnamefont
  {Parto}}, \bibinfo {author} {\bibfnamefont {Y.~G.~N.}\ \bibnamefont {Liu}},
  \bibinfo {author} {\bibfnamefont {B.}~\bibnamefont {Bahari}}, \bibinfo
  {author} {\bibfnamefont {M.}~\bibnamefont {Khajavikhan}}, \ and\ \bibinfo
  {author} {\bibfnamefont {D.~N.}\ \bibnamefont {Christodoulides}},\ }\bibfield
   {title} {\enquote {\bibinfo {title} {Non-hermitian and topological
  photonics: optics at an exceptional point},}\ }\href {\doibase
  doi:10.1515/nanoph-2020-0434} {\bibfield  {journal} {\bibinfo  {journal}
  {Nanophotonics}\ }\textbf {\bibinfo {volume} {10}},\ \bibinfo {pages} {403}
  (\bibinfo {year} {2021})}\BibitemShut {NoStop}%
\bibitem [{\citenamefont {Ding}\ \emph {et~al.}(2022)\citenamefont {Ding},
  \citenamefont {Fang},\ and\ \citenamefont {Ma}}]{Ding:2022juv}%
  \BibitemOpen
  \bibfield  {author} {\bibinfo {author} {\bibfnamefont {K.}~\bibnamefont
  {Ding}}, \bibinfo {author} {\bibfnamefont {C.}~\bibnamefont {Fang}}, \ and\
  \bibinfo {author} {\bibfnamefont {G.}~\bibnamefont {Ma}},\ }\bibfield
  {title} {\enquote {\bibinfo {title} {{Non-Hermitian topology and
  exceptional-point geometries}},}\ }\href {\doibase
  10.1038/s42254-022-00516-5} {\bibfield  {journal} {\bibinfo  {journal}
  {Nature Rev. Phys.}\ }\textbf {\bibinfo {volume} {4}},\ \bibinfo {pages}
  {745} (\bibinfo {year} {2022})},\ \Eprint {http://arxiv.org/abs/2204.11601}
  {arXiv:2204.11601 [quant-ph]} \BibitemShut {NoStop}%
\bibitem [{\citenamefont {Dietz}\ \emph {et~al.}(2007)\citenamefont {Dietz},
  \citenamefont {Friedrich}, \citenamefont {Metz}, \citenamefont {Miski-Oglu},
  \citenamefont {Richter}, \citenamefont {Sch\"afer},\ and\ \citenamefont
  {Stafford}}]{PhysRevE.75.027201}%
  \BibitemOpen
  \bibfield  {author} {\bibinfo {author} {\bibfnamefont {B.}~\bibnamefont
  {Dietz}}, \bibinfo {author} {\bibfnamefont {T.}~\bibnamefont {Friedrich}},
  \bibinfo {author} {\bibfnamefont {J.}~\bibnamefont {Metz}}, \bibinfo {author}
  {\bibfnamefont {M.}~\bibnamefont {Miski-Oglu}}, \bibinfo {author}
  {\bibfnamefont {A.}~\bibnamefont {Richter}}, \bibinfo {author} {\bibfnamefont
  {F.}~\bibnamefont {Sch\"afer}}, \ and\ \bibinfo {author} {\bibfnamefont
  {C.~A.}\ \bibnamefont {Stafford}},\ }\bibfield  {title} {\enquote {\bibinfo
  {title} {Rabi oscillations at exceptional points in microwave billiards},}\
  }\href {\doibase 10.1103/PhysRevE.75.027201} {\bibfield  {journal} {\bibinfo
  {journal} {Phys. Rev. E}\ }\textbf {\bibinfo {volume} {75}},\ \bibinfo
  {pages} {027201} (\bibinfo {year} {2007})},\ \Eprint
  {http://arxiv.org/abs/cond-mat/0612547} {arXiv:cond-mat/0612547} \BibitemShut
  {NoStop}%
\bibitem [{\citenamefont {Verstraete}\ \emph {et~al.}(2008)\citenamefont
  {Verstraete}, \citenamefont {Murg},\ and\ \citenamefont
  {Cirac}}]{Verstraete01032008}%
  \BibitemOpen
  \bibfield  {author} {\bibinfo {author} {\bibfnamefont {F.}~\bibnamefont
  {Verstraete}}, \bibinfo {author} {\bibfnamefont {V.}~\bibnamefont {Murg}}, \
  and\ \bibinfo {author} {\bibfnamefont {J.}~\bibnamefont {Cirac}},\ }\bibfield
   {title} {\enquote {\bibinfo {title} {Matrix product states, projected
  entangled pair states, and variational renormalization group methods for
  quantum spin systems},}\ }\href {\doibase 10.1080/14789940801912366}
  {\bibfield  {journal} {\bibinfo  {journal} {Advances in Physics}\ }\textbf
  {\bibinfo {volume} {57}},\ \bibinfo {pages} {143} (\bibinfo {year} {2008})},\
  \Eprint {http://arxiv.org/abs/0907.2796} {arXiv:0907.2796 [quant-ph]}
  \BibitemShut {NoStop}%
\bibitem [{\citenamefont {Heiss}(2010)}]{Heiss2010}%
  \BibitemOpen
  \bibfield  {author} {\bibinfo {author} {\bibfnamefont {W.~D.}\ \bibnamefont
  {Heiss}},\ }\bibfield  {title} {\enquote {\bibinfo {title} {Time behaviour
  near to spectral singularities},}\ }\href {\doibase
  10.1140/epjd/e2010-00243-0} {\bibfield  {journal} {\bibinfo  {journal} {The
  European Physical Journal D}\ }\textbf {\bibinfo {volume} {60}},\ \bibinfo
  {pages} {257} (\bibinfo {year} {2010})},\ \Eprint
  {http://arxiv.org/abs/1009.5780} {arXiv:1009.5780 [quant-ph]} \BibitemShut
  {NoStop}%
\bibitem [{\citenamefont {Jaramillo}\ \emph {et~al.}(2021)\citenamefont
  {Jaramillo}, \citenamefont {Panosso~Macedo},\ and\ \citenamefont
  {Al~Sheikh}}]{Jaramillo:2020tuu}%
  \BibitemOpen
  \bibfield  {author} {\bibinfo {author} {\bibfnamefont {J.~L.}\ \bibnamefont
  {Jaramillo}}, \bibinfo {author} {\bibfnamefont {R.}~\bibnamefont
  {Panosso~Macedo}}, \ and\ \bibinfo {author} {\bibfnamefont {L.}~\bibnamefont
  {Al~Sheikh}},\ }\bibfield  {title} {\enquote {\bibinfo {title}
  {{Pseudospectrum and Black Hole Quasinormal Mode Instability}},}\ }\href
  {\doibase 10.1103/PhysRevX.11.031003} {\bibfield  {journal} {\bibinfo
  {journal} {Phys. Rev. X}\ }\textbf {\bibinfo {volume} {11}},\ \bibinfo
  {pages} {031003} (\bibinfo {year} {2021})},\ \Eprint
  {http://arxiv.org/abs/2004.06434} {arXiv:2004.06434 [gr-qc]} \BibitemShut
  {NoStop}%
\bibitem [{\citenamefont {Jaramillo}(2022)}]{Jaramillo:2022kuv}%
  \BibitemOpen
  \bibfield  {author} {\bibinfo {author} {\bibfnamefont {J.~L.}\ \bibnamefont
  {Jaramillo}},\ }\bibfield  {title} {\enquote {\bibinfo {title}
  {{Pseudospectrum and binary black hole merger transients}},}\ }\href
  {\doibase 10.1088/1361-6382/ac8ddc} {\bibfield  {journal} {\bibinfo
  {journal} {Class. Quant. Grav.}\ }\textbf {\bibinfo {volume} {39}},\ \bibinfo
  {pages} {217002} (\bibinfo {year} {2022})},\ \Eprint
  {http://arxiv.org/abs/2206.08025} {arXiv:2206.08025 [gr-qc]} \BibitemShut
  {NoStop}%
\bibitem [{\citenamefont {Motohashi}(2025)}]{Motohashi:2024fwt}%
  \BibitemOpen
  \bibfield  {author} {\bibinfo {author} {\bibfnamefont {H.}~\bibnamefont
  {Motohashi}},\ }\bibfield  {title} {\enquote {\bibinfo {title} {{Resonant
  Excitation of Quasinormal Modes of Black Holes}},}\ }\href {\doibase
  10.1103/PhysRevLett.134.141401} {\bibfield  {journal} {\bibinfo  {journal}
  {Phys. Rev. Lett.}\ }\textbf {\bibinfo {volume} {134}},\ \bibinfo {pages}
  {141401} (\bibinfo {year} {2025})},\ \Eprint
  {http://arxiv.org/abs/2407.15191} {arXiv:2407.15191 [gr-qc]} \BibitemShut
  {NoStop}%
\bibitem [{\citenamefont {Cavalcante}\ \emph
  {et~al.}(2024{\natexlab{a}})\citenamefont {Cavalcante}, \citenamefont
  {Richartz},\ and\ \citenamefont {da~Cunha}}]{Cavalcante:2024swt}%
  \BibitemOpen
  \bibfield  {author} {\bibinfo {author} {\bibfnamefont {J.~P.}\ \bibnamefont
  {Cavalcante}}, \bibinfo {author} {\bibfnamefont {M.}~\bibnamefont
  {Richartz}}, \ and\ \bibinfo {author} {\bibfnamefont {B.~C.}\ \bibnamefont
  {da~Cunha}},\ }\bibfield  {title} {\enquote {\bibinfo {title} {{Exceptional
  Point and Hysteresis in Perturbations of Kerr Black Holes}},}\ }\href
  {\doibase 10.1103/PhysRevLett.133.261401} {\bibfield  {journal} {\bibinfo
  {journal} {Phys. Rev. Lett.}\ }\textbf {\bibinfo {volume} {133}},\ \bibinfo
  {pages} {261401} (\bibinfo {year} {2024}{\natexlab{a}})},\ \Eprint
  {http://arxiv.org/abs/2407.20850} {arXiv:2407.20850 [gr-qc]} \BibitemShut
  {NoStop}%
\bibitem [{\citenamefont {Cavalcante}\ \emph
  {et~al.}(2024{\natexlab{b}})\citenamefont {Cavalcante}, \citenamefont
  {Richartz},\ and\ \citenamefont {da~Cunha}}]{Cavalcante:2024kmy}%
  \BibitemOpen
  \bibfield  {author} {\bibinfo {author} {\bibfnamefont {J.~P.}\ \bibnamefont
  {Cavalcante}}, \bibinfo {author} {\bibfnamefont {M.}~\bibnamefont
  {Richartz}}, \ and\ \bibinfo {author} {\bibfnamefont {B.~C.}\ \bibnamefont
  {da~Cunha}},\ }\bibfield  {title} {\enquote {\bibinfo {title} {{Massive
  scalar perturbations in Kerr black holes: Near extremal analysis}},}\ }\href
  {\doibase 10.1103/PhysRevD.110.124064} {\bibfield  {journal} {\bibinfo
  {journal} {Phys. Rev. D}\ }\textbf {\bibinfo {volume} {110}},\ \bibinfo
  {pages} {124064} (\bibinfo {year} {2024}{\natexlab{b}})},\ \Eprint
  {http://arxiv.org/abs/2408.13964} {arXiv:2408.13964 [gr-qc]} \BibitemShut
  {NoStop}%
\bibitem [{\citenamefont {{Andersson}}\ \emph {et~al.}(1996)\citenamefont
  {{Andersson}}, \citenamefont {{Kojima}},\ and\ \citenamefont
  {{Kokkotas}}}]{1996ApJ...462..855A}%
  \BibitemOpen
  \bibfield  {author} {\bibinfo {author} {\bibfnamefont {N.}~\bibnamefont
  {{Andersson}}}, \bibinfo {author} {\bibfnamefont {Y.}~\bibnamefont
  {{Kojima}}}, \ and\ \bibinfo {author} {\bibfnamefont {K.~D.}\ \bibnamefont
  {{Kokkotas}}},\ }\bibfield  {title} {\enquote {\bibinfo {title} {{On the
  Oscillation Spectra of Ultracompact Stars: an Extensive Survey of
  Gravitational-Wave Modes}},}\ }\href {\doibase 10.1086/177199} {\bibfield
  {journal} {\bibinfo  {journal} {\apj}\ }\textbf {\bibinfo {volume} {462}},\
  \bibinfo {pages} {855} (\bibinfo {year} {1996})},\ \Eprint
  {http://arxiv.org/abs/gr-qc/9512048} {arXiv:gr-qc/9512048 [gr-qc]}
  \BibitemShut {NoStop}%
\bibitem [{\citenamefont {Andersson}\ and\ \citenamefont
  {Kokkotas}(1998)}]{Andersson:1997eq}%
  \BibitemOpen
  \bibfield  {author} {\bibinfo {author} {\bibfnamefont {N.}~\bibnamefont
  {Andersson}}\ and\ \bibinfo {author} {\bibfnamefont {K.~D.}\ \bibnamefont
  {Kokkotas}},\ }\bibfield  {title} {\enquote {\bibinfo {title} {{Pulsation
  modes for increasingly relativistic polytropes}},}\ }\href {\doibase
  10.1046/j.1365-8711.1998.01541.x} {\bibfield  {journal} {\bibinfo  {journal}
  {Mon. Not. Roy. Astron. Soc.}\ }\textbf {\bibinfo {volume} {297}},\ \bibinfo
  {pages} {493} (\bibinfo {year} {1998})},\ \Eprint
  {http://arxiv.org/abs/gr-qc/9706010} {arXiv:gr-qc/9706010} \BibitemShut
  {NoStop}%
\bibitem [{\citenamefont {Gondek}\ \emph {et~al.}(1997)\citenamefont {Gondek},
  \citenamefont {Haensel},\ and\ \citenamefont {Zdunik}}]{Gondek:1997fd}%
  \BibitemOpen
  \bibfield  {author} {\bibinfo {author} {\bibfnamefont {D.}~\bibnamefont
  {Gondek}}, \bibinfo {author} {\bibfnamefont {P.}~\bibnamefont {Haensel}}, \
  and\ \bibinfo {author} {\bibfnamefont {J.~L.}\ \bibnamefont {Zdunik}},\
  }\bibfield  {title} {\enquote {\bibinfo {title} {{Radial pulsations and
  stability of protoneutron stars}},}\ }\href@noop {} {\bibfield  {journal}
  {\bibinfo  {journal} {Astron. Astrophys.}\ }\textbf {\bibinfo {volume}
  {325}},\ \bibinfo {pages} {217} (\bibinfo {year} {1997})},\ \Eprint
  {http://arxiv.org/abs/astro-ph/9705157} {arXiv:astro-ph/9705157} \BibitemShut
  {NoStop}%
\bibitem [{\citenamefont {{Aizenman}}\ \emph {et~al.}(1977)\citenamefont
  {{Aizenman}}, \citenamefont {{Smeyers}},\ and\ \citenamefont
  {{Weigert}}}]{1977A&A....58...41A}%
  \BibitemOpen
  \bibfield  {author} {\bibinfo {author} {\bibfnamefont {M.}~\bibnamefont
  {{Aizenman}}}, \bibinfo {author} {\bibfnamefont {P.}~\bibnamefont
  {{Smeyers}}}, \ and\ \bibinfo {author} {\bibfnamefont {A.}~\bibnamefont
  {{Weigert}}},\ }\bibfield  {title} {\enquote {\bibinfo {title} {{Avoided
  Crossing of Modes of Non-radial Stellar Oscillations}},}\ }\href@noop {}
  {\bibfield  {journal} {\bibinfo  {journal} {Astron. Astrophys.}\ }\textbf
  {\bibinfo {volume} {58}},\ \bibinfo {pages} {41} (\bibinfo {year}
  {1977})}\BibitemShut {NoStop}%
\bibitem [{\citenamefont {Sotani}\ and\ \citenamefont
  {Takiwaki}(2020)}]{Sotani:2020eva}%
  \BibitemOpen
  \bibfield  {author} {\bibinfo {author} {\bibfnamefont {H.}~\bibnamefont
  {Sotani}}\ and\ \bibinfo {author} {\bibfnamefont {T.}~\bibnamefont
  {Takiwaki}},\ }\bibfield  {title} {\enquote {\bibinfo {title} {{Avoided
  crossing in gravitational wave spectra from protoneutron star}},}\ }\href
  {\doibase 10.1093/mnras/staa2597} {\bibfield  {journal} {\bibinfo  {journal}
  {Mon. Not. Roy. Astron. Soc.}\ }\textbf {\bibinfo {volume} {498}},\ \bibinfo
  {pages} {3503} (\bibinfo {year} {2020})},\ \Eprint
  {http://arxiv.org/abs/2008.00419} {arXiv:2008.00419 [astro-ph.HE]}
  \BibitemShut {NoStop}%
\bibitem [{\citenamefont {Berti}\ \emph {et~al.}(2025)\citenamefont {Berti}
  \emph {et~al.}}]{Berti:2025hly}%
  \BibitemOpen
  \bibfield  {author} {\bibinfo {author} {\bibfnamefont {E.}~\bibnamefont
  {Berti}} \emph {et~al.},\ }\bibfield  {title} {\enquote {\bibinfo {title}
  {{Black hole spectroscopy: from theory to experiment}},}\ }\href@noop {} {\
  (\bibinfo {year} {2025})},\ \Eprint {http://arxiv.org/abs/2505.23895}
  {arXiv:2505.23895 [gr-qc]} \BibitemShut {NoStop}%
\bibitem [{\citenamefont {Oshita}\ \emph {et~al.}(2025)\citenamefont {Oshita},
  \citenamefont {Berti},\ and\ \citenamefont {Cardoso}}]{Oshita:2025ibu}%
  \BibitemOpen
  \bibfield  {author} {\bibinfo {author} {\bibfnamefont {N.}~\bibnamefont
  {Oshita}}, \bibinfo {author} {\bibfnamefont {E.}~\bibnamefont {Berti}}, \
  and\ \bibinfo {author} {\bibfnamefont {V.}~\bibnamefont {Cardoso}},\
  }\bibfield  {title} {\enquote {\bibinfo {title} {{Unstable Chords and
  Destructive Resonant Excitation of Black Hole Quasinormal Modes}},}\ }\href
  {\doibase 10.1103/ht2n-vvvh} {\bibfield  {journal} {\bibinfo  {journal}
  {Phys. Rev. Lett.}\ }\textbf {\bibinfo {volume} {135}},\ \bibinfo {pages}
  {031401} (\bibinfo {year} {2025})},\ \Eprint
  {http://arxiv.org/abs/2503.21276} {arXiv:2503.21276 [gr-qc]} \BibitemShut
  {NoStop}%
\bibitem [{\citenamefont {Lo}\ \emph {et~al.}(2025)\citenamefont {Lo},
  \citenamefont {Sabani},\ and\ \citenamefont {Cardoso}}]{Lo:2025njp}%
  \BibitemOpen
  \bibfield  {author} {\bibinfo {author} {\bibfnamefont {R.~K.~L.}\
  \bibnamefont {Lo}}, \bibinfo {author} {\bibfnamefont {L.}~\bibnamefont
  {Sabani}}, \ and\ \bibinfo {author} {\bibfnamefont {V.}~\bibnamefont
  {Cardoso}},\ }\bibfield  {title} {\enquote {\bibinfo {title} {{Quasinormal
  modes and excitation factors of Kerr black holes}},}\ }\href {\doibase
  10.1103/PhysRevD.111.124002} {\bibfield  {journal} {\bibinfo  {journal}
  {Phys. Rev. D}\ }\textbf {\bibinfo {volume} {111}},\ \bibinfo {pages}
  {124002} (\bibinfo {year} {2025})},\ \Eprint
  {http://arxiv.org/abs/2504.00084} {arXiv:2504.00084 [gr-qc]} \BibitemShut
  {NoStop}%
\bibitem [{\citenamefont {Yang}\ \emph {et~al.}(2025)\citenamefont {Yang},
  \citenamefont {Berti},\ and\ \citenamefont {Franchini}}]{Yang:2025dbn}%
  \BibitemOpen
  \bibfield  {author} {\bibinfo {author} {\bibfnamefont {Y.}~\bibnamefont
  {Yang}}, \bibinfo {author} {\bibfnamefont {E.}~\bibnamefont {Berti}}, \ and\
  \bibinfo {author} {\bibfnamefont {N.}~\bibnamefont {Franchini}},\ }\bibfield
  {title} {\enquote {\bibinfo {title} {{Black Hole Quasinormal Mode
  Resonances}},}\ }\href {\doibase 10.1103/hfv8-n444} {\bibfield  {journal}
  {\bibinfo  {journal} {Phys. Rev. Lett.}\ }\textbf {\bibinfo {volume} {135}},\
  \bibinfo {pages} {201401} (\bibinfo {year} {2025})},\ \Eprint
  {http://arxiv.org/abs/2504.06072} {arXiv:2504.06072 [gr-qc]} \BibitemShut
  {NoStop}%
\bibitem [{\citenamefont {Takahashi}\ \emph {et~al.}(2025)\citenamefont
  {Takahashi}, \citenamefont {Motohashi},\ and\ \citenamefont
  {Takahashi}}]{Takahashi:2025uwo}%
  \BibitemOpen
  \bibfield  {author} {\bibinfo {author} {\bibfnamefont {T.}~\bibnamefont
  {Takahashi}}, \bibinfo {author} {\bibfnamefont {H.}~\bibnamefont
  {Motohashi}}, \ and\ \bibinfo {author} {\bibfnamefont {K.}~\bibnamefont
  {Takahashi}},\ }\bibfield  {title} {\enquote {\bibinfo {title} {{Resonance of
  black hole quasinormal modes in coupled systems}},}\ }\href {\doibase
  10.1103/59tb-2x9r} {\bibfield  {journal} {\bibinfo  {journal} {Phys. Rev. D}\
  }\textbf {\bibinfo {volume} {112}},\ \bibinfo {pages} {064006} (\bibinfo
  {year} {2025})},\ \Eprint {http://arxiv.org/abs/2505.03883} {arXiv:2505.03883
  [gr-qc]} \BibitemShut {NoStop}%
\bibitem [{\citenamefont {Chen}\ \emph {et~al.}(2025)\citenamefont {Chen},
  \citenamefont {Jing}, \citenamefont {Cao},\ and\ \citenamefont
  {Wang}}]{Chen:2025sbz}%
  \BibitemOpen
  \bibfield  {author} {\bibinfo {author} {\bibfnamefont {C.}~\bibnamefont
  {Chen}}, \bibinfo {author} {\bibfnamefont {J.}~\bibnamefont {Jing}}, \bibinfo
  {author} {\bibfnamefont {Z.}~\bibnamefont {Cao}}, \ and\ \bibinfo {author}
  {\bibfnamefont {M.}~\bibnamefont {Wang}},\ }\bibfield  {title} {\enquote
  {\bibinfo {title} {{Complete quasinormal modes of type-D black holes}},}\
  }\href {\doibase 10.1103/f8m8-vr4l} {\bibfield  {journal} {\bibinfo
  {journal} {Phys. Rev. D}\ }\textbf {\bibinfo {volume} {112}},\ \bibinfo
  {pages} {103036} (\bibinfo {year} {2025})},\ \Eprint
  {http://arxiv.org/abs/2506.14635} {arXiv:2506.14635 [gr-qc]} \BibitemShut
  {NoStop}%
\bibitem [{\citenamefont {Kubota}\ and\ \citenamefont
  {Motohashi}(2026)}]{Kubota:2025hjk}%
  \BibitemOpen
  \bibfield  {author} {\bibinfo {author} {\bibfnamefont {K.-i.}\ \bibnamefont
  {Kubota}}\ and\ \bibinfo {author} {\bibfnamefont {H.}~\bibnamefont
  {Motohashi}},\ }\bibfield  {title} {\enquote {\bibinfo {title} {{Resonance in
  black hole ringdown: Benchmarking quasinormal mode excitation and
  extraction}},}\ }\href {\doibase 10.1103/pwyd-nv6v} {\bibfield  {journal}
  {\bibinfo  {journal} {Phys. Rev. D}\ }\textbf {\bibinfo {volume} {113}},\
  \bibinfo {pages} {043053} (\bibinfo {year} {2026})},\ \Eprint
  {http://arxiv.org/abs/2509.06411} {arXiv:2509.06411 [gr-qc]} \BibitemShut
  {NoStop}%
\bibitem [{\citenamefont {Cavalcante}\ \emph {et~al.}(2025)\citenamefont
  {Cavalcante}, \citenamefont {Richartz},\ and\ \citenamefont
  {da~Cunha}}]{Cavalcante:2025abr}%
  \BibitemOpen
  \bibfield  {author} {\bibinfo {author} {\bibfnamefont {J.~P.}\ \bibnamefont
  {Cavalcante}}, \bibinfo {author} {\bibfnamefont {M.}~\bibnamefont
  {Richartz}}, \ and\ \bibinfo {author} {\bibfnamefont {B.~C.}\ \bibnamefont
  {da~Cunha}},\ }\bibfield  {title} {\enquote {\bibinfo {title} {{Ergodic
  Hysteresis of the Kerr black hole spectrum}},}\ }\href@noop {} {\  (\bibinfo
  {year} {2025})},\ \Eprint {http://arxiv.org/abs/2511.16640} {arXiv:2511.16640
  [gr-qc]} \BibitemShut {NoStop}%
\bibitem [{\citenamefont {Cao}\ \emph {et~al.}(2025)\citenamefont {Cao},
  \citenamefont {Ji}, \citenamefont {Wu},\ and\ \citenamefont
  {Zhou}}]{Cao:2025afs}%
  \BibitemOpen
  \bibfield  {author} {\bibinfo {author} {\bibfnamefont {L.-M.}\ \bibnamefont
  {Cao}}, \bibinfo {author} {\bibfnamefont {M.-F.}\ \bibnamefont {Ji}},
  \bibinfo {author} {\bibfnamefont {L.-B.}\ \bibnamefont {Wu}}, \ and\ \bibinfo
  {author} {\bibfnamefont {Y.-S.}\ \bibnamefont {Zhou}},\ }\bibfield  {title}
  {\enquote {\bibinfo {title} {{Exceptional line and pseudospectrum in black
  hole spectroscopy}},}\ }\href@noop {} {\  (\bibinfo {year} {2025})},\ \Eprint
  {http://arxiv.org/abs/2511.17067} {arXiv:2511.17067 [gr-qc]} \BibitemShut
  {NoStop}%
\bibitem [{\citenamefont {Leaver}(1986)}]{Leaver:1986gd}%
  \BibitemOpen
  \bibfield  {author} {\bibinfo {author} {\bibfnamefont {E.~W.}\ \bibnamefont
  {Leaver}},\ }\bibfield  {title} {\enquote {\bibinfo {title} {{Spectral
  decomposition of the perturbation response of the Schwarzschild geometry}},}\
  }\href {\doibase 10.1103/PhysRevD.34.384} {\bibfield  {journal} {\bibinfo
  {journal} {Phys. Rev. D}\ }\textbf {\bibinfo {volume} {34}},\ \bibinfo
  {pages} {384} (\bibinfo {year} {1986})}\BibitemShut {NoStop}%
\bibitem [{\citenamefont {Onozawa}(1997)}]{Onozawa:1996ux}%
  \BibitemOpen
  \bibfield  {author} {\bibinfo {author} {\bibfnamefont {H.}~\bibnamefont
  {Onozawa}},\ }\bibfield  {title} {\enquote {\bibinfo {title} {{A Detailed
  study of quasinormal frequencies of the Kerr black hole}},}\ }\href {\doibase
  10.1103/PhysRevD.55.3593} {\bibfield  {journal} {\bibinfo  {journal} {Phys.
  Rev. D}\ }\textbf {\bibinfo {volume} {55}},\ \bibinfo {pages} {3593}
  (\bibinfo {year} {1997})},\ \Eprint {http://arxiv.org/abs/gr-qc/9610048}
  {arXiv:gr-qc/9610048} \BibitemShut {NoStop}%
\bibitem [{\citenamefont {Yang}\ \emph {et~al.}(2013)\citenamefont {Yang},
  \citenamefont {Zimmerman}, \citenamefont {Zengino\u{g}lu}, \citenamefont
  {Zhang}, \citenamefont {Berti},\ and\ \citenamefont {Chen}}]{Yang:2013uba}%
  \BibitemOpen
  \bibfield  {author} {\bibinfo {author} {\bibfnamefont {H.}~\bibnamefont
  {Yang}}, \bibinfo {author} {\bibfnamefont {A.}~\bibnamefont {Zimmerman}},
  \bibinfo {author} {\bibfnamefont {A.}~\bibnamefont {Zengino\u{g}lu}},
  \bibinfo {author} {\bibfnamefont {F.}~\bibnamefont {Zhang}}, \bibinfo
  {author} {\bibfnamefont {E.}~\bibnamefont {Berti}}, \ and\ \bibinfo {author}
  {\bibfnamefont {Y.}~\bibnamefont {Chen}},\ }\bibfield  {title} {\enquote
  {\bibinfo {title} {{Quasinormal modes of nearly extremal Kerr spacetimes:
  spectrum bifurcation and power-law ringdown}},}\ }\href {\doibase
  10.1103/PhysRevD.88.044047} {\bibfield  {journal} {\bibinfo  {journal} {Phys.
  Rev. D}\ }\textbf {\bibinfo {volume} {88}},\ \bibinfo {pages} {044047}
  (\bibinfo {year} {2013})},\ \Eprint {http://arxiv.org/abs/1307.8086}
  {arXiv:1307.8086 [gr-qc]} \BibitemShut {NoStop}%
\bibitem [{\citenamefont {Cook}\ and\ \citenamefont
  {Zalutskiy}(2014)}]{Cook:2014cta}%
  \BibitemOpen
  \bibfield  {author} {\bibinfo {author} {\bibfnamefont {G.~B.}\ \bibnamefont
  {Cook}}\ and\ \bibinfo {author} {\bibfnamefont {M.}~\bibnamefont
  {Zalutskiy}},\ }\bibfield  {title} {\enquote {\bibinfo {title}
  {{Gravitational perturbations of the Kerr geometry: High-accuracy study}},}\
  }\href {\doibase 10.1103/PhysRevD.90.124021} {\bibfield  {journal} {\bibinfo
  {journal} {Phys. Rev. D}\ }\textbf {\bibinfo {volume} {90}},\ \bibinfo
  {pages} {124021} (\bibinfo {year} {2014})},\ \Eprint
  {http://arxiv.org/abs/1410.7698} {arXiv:1410.7698 [gr-qc]} \BibitemShut
  {NoStop}%
\bibitem [{\citenamefont {Jansen}(2017)}]{Jansen:2017oag}%
  \BibitemOpen
  \bibfield  {author} {\bibinfo {author} {\bibfnamefont {A.}~\bibnamefont
  {Jansen}},\ }\bibfield  {title} {\enquote {\bibinfo {title} {{Overdamped
  modes in Schwarzschild-de Sitter and a Mathematica package for the numerical
  computation of quasinormal modes}},}\ }\href {\doibase
  10.1140/epjp/i2017-11825-9} {\bibfield  {journal} {\bibinfo  {journal} {Eur.
  Phys. J. Plus}\ }\textbf {\bibinfo {volume} {132}},\ \bibinfo {pages} {546}
  (\bibinfo {year} {2017})},\ \Eprint {http://arxiv.org/abs/1709.09178}
  {arXiv:1709.09178 [gr-qc]} \BibitemShut {NoStop}%
\bibitem [{\citenamefont {Dias}\ \emph
  {et~al.}(2022{\natexlab{a}})\citenamefont {Dias}, \citenamefont {Godazgar},
  \citenamefont {Santos}, \citenamefont {Carullo}, \citenamefont {Del~Pozzo},\
  and\ \citenamefont {Laghi}}]{Dias:2021yju}%
  \BibitemOpen
  \bibfield  {author} {\bibinfo {author} {\bibfnamefont {O.~J.~C.}\
  \bibnamefont {Dias}}, \bibinfo {author} {\bibfnamefont {M.}~\bibnamefont
  {Godazgar}}, \bibinfo {author} {\bibfnamefont {J.~E.}\ \bibnamefont
  {Santos}}, \bibinfo {author} {\bibfnamefont {G.}~\bibnamefont {Carullo}},
  \bibinfo {author} {\bibfnamefont {W.}~\bibnamefont {Del~Pozzo}}, \ and\
  \bibinfo {author} {\bibfnamefont {D.}~\bibnamefont {Laghi}},\ }\bibfield
  {title} {\enquote {\bibinfo {title} {{Eigenvalue repulsions in the
  quasinormal spectra of the Kerr-Newman black hole}},}\ }\href {\doibase
  10.1103/PhysRevD.105.084044} {\bibfield  {journal} {\bibinfo  {journal}
  {Phys. Rev. D}\ }\textbf {\bibinfo {volume} {105}},\ \bibinfo {pages}
  {084044} (\bibinfo {year} {2022}{\natexlab{a}})},\ \Eprint
  {http://arxiv.org/abs/2109.13949} {arXiv:2109.13949 [gr-qc]} \BibitemShut
  {NoStop}%
\bibitem [{\citenamefont {Davey}\ \emph {et~al.}(2022)\citenamefont {Davey},
  \citenamefont {Dias}, \citenamefont {Rodgers},\ and\ \citenamefont
  {Santos}}]{Davey:2022vyx}%
  \BibitemOpen
  \bibfield  {author} {\bibinfo {author} {\bibfnamefont {A.}~\bibnamefont
  {Davey}}, \bibinfo {author} {\bibfnamefont {O.~J.~C.}\ \bibnamefont {Dias}},
  \bibinfo {author} {\bibfnamefont {P.}~\bibnamefont {Rodgers}}, \ and\
  \bibinfo {author} {\bibfnamefont {J.~E.}\ \bibnamefont {Santos}},\ }\bibfield
   {title} {\enquote {\bibinfo {title} {{Strong Cosmic Censorship and
  eigenvalue repulsions for rotating de Sitter black holes in
  higher-dimensions}},}\ }\href {\doibase 10.1007/JHEP07(2022)086} {\bibfield
  {journal} {\bibinfo  {journal} {JHEP}\ }\textbf {\bibinfo {volume} {07}},\
  \bibinfo {pages} {086} (\bibinfo {year} {2022})},\ \Eprint
  {http://arxiv.org/abs/2203.13830} {arXiv:2203.13830 [gr-qc]} \BibitemShut
  {NoStop}%
\bibitem [{\citenamefont {Dias}\ \emph
  {et~al.}(2022{\natexlab{b}})\citenamefont {Dias}, \citenamefont {Godazgar},\
  and\ \citenamefont {Santos}}]{Dias:2022oqm}%
  \BibitemOpen
  \bibfield  {author} {\bibinfo {author} {\bibfnamefont {O.~J.~C.}\
  \bibnamefont {Dias}}, \bibinfo {author} {\bibfnamefont {M.}~\bibnamefont
  {Godazgar}}, \ and\ \bibinfo {author} {\bibfnamefont {J.~E.}\ \bibnamefont
  {Santos}},\ }\bibfield  {title} {\enquote {\bibinfo {title} {{Eigenvalue
  repulsions and quasinormal mode spectra of Kerr-Newman: an extended
  study}},}\ }\href {\doibase 10.1007/JHEP07(2022)076} {\bibfield  {journal}
  {\bibinfo  {journal} {JHEP}\ }\textbf {\bibinfo {volume} {07}},\ \bibinfo
  {pages} {076} (\bibinfo {year} {2022}{\natexlab{b}})},\ \Eprint
  {http://arxiv.org/abs/2205.13072} {arXiv:2205.13072 [gr-qc]} \BibitemShut
  {NoStop}%
\bibitem [{\citenamefont {Kinoshita}\ \emph {et~al.}(2024)\citenamefont
  {Kinoshita}, \citenamefont {Kozuka}, \citenamefont {Murata},\ and\
  \citenamefont {Sugawara}}]{Kinoshita:2023iad}%
  \BibitemOpen
  \bibfield  {author} {\bibinfo {author} {\bibfnamefont {S.}~\bibnamefont
  {Kinoshita}}, \bibinfo {author} {\bibfnamefont {T.}~\bibnamefont {Kozuka}},
  \bibinfo {author} {\bibfnamefont {K.}~\bibnamefont {Murata}}, \ and\ \bibinfo
  {author} {\bibfnamefont {K.}~\bibnamefont {Sugawara}},\ }\bibfield  {title}
  {\enquote {\bibinfo {title} {{Quasinormal mode spectrum of the AdS black hole
  with the Robin boundary condition}},}\ }\href {\doibase
  10.1088/1361-6382/ad1cbe} {\bibfield  {journal} {\bibinfo  {journal} {Class.
  Quant. Grav.}\ }\textbf {\bibinfo {volume} {41}},\ \bibinfo {pages} {055010}
  (\bibinfo {year} {2024})},\ \Eprint {http://arxiv.org/abs/2305.17942}
  {arXiv:2305.17942 [gr-qc]} \BibitemShut {NoStop}%
\bibitem [{\citenamefont {Chandrasekhar}(1985)}]{Chandrasekhar:1985kt}%
  \BibitemOpen
  \bibfield  {author} {\bibinfo {author} {\bibfnamefont {S.}~\bibnamefont
  {Chandrasekhar}},\ }\href@noop {} {\emph {\bibinfo {title} {{The mathematical
  theory of black holes}}}}\ (\bibinfo  {publisher} {Oxford University Press},\
  \bibinfo {year} {1985})\BibitemShut {NoStop}%
\bibitem [{Note1()}]{Note1}%
  \BibitemOpen
  \bibinfo {note} {We omit the angular indices $(\ell ,m)$ for
  simplicity.}\BibitemShut {Stop}%
\bibitem [{\citenamefont {Cheung}\ \emph {et~al.}(2022)\citenamefont {Cheung},
  \citenamefont {Destounis}, \citenamefont {Macedo}, \citenamefont {Berti},\
  and\ \citenamefont {Cardoso}}]{Cheung:2021bol}%
  \BibitemOpen
  \bibfield  {author} {\bibinfo {author} {\bibfnamefont {M.~H.-Y.}\
  \bibnamefont {Cheung}}, \bibinfo {author} {\bibfnamefont {K.}~\bibnamefont
  {Destounis}}, \bibinfo {author} {\bibfnamefont {R.~P.}\ \bibnamefont
  {Macedo}}, \bibinfo {author} {\bibfnamefont {E.}~\bibnamefont {Berti}}, \
  and\ \bibinfo {author} {\bibfnamefont {V.}~\bibnamefont {Cardoso}},\
  }\bibfield  {title} {\enquote {\bibinfo {title} {{Destabilizing the
  Fundamental Mode of Black Holes: The Elephant and the Flea}},}\ }\href
  {\doibase 10.1103/PhysRevLett.128.111103} {\bibfield  {journal} {\bibinfo
  {journal} {Phys. Rev. Lett.}\ }\textbf {\bibinfo {volume} {128}},\ \bibinfo
  {pages} {111103} (\bibinfo {year} {2022})},\ \Eprint
  {http://arxiv.org/abs/2111.05415} {arXiv:2111.05415 [gr-qc]} \BibitemShut
  {NoStop}%
\bibitem [{\citenamefont {Zenginoglu}(2011)}]{Zenginoglu:2011jz}%
  \BibitemOpen
  \bibfield  {author} {\bibinfo {author} {\bibfnamefont {A.}~\bibnamefont
  {Zenginoglu}},\ }\bibfield  {title} {\enquote {\bibinfo {title} {{A Geometric
  framework for black hole perturbations}},}\ }\href {\doibase
  10.1103/PhysRevD.83.127502} {\bibfield  {journal} {\bibinfo  {journal} {Phys.
  Rev. D}\ }\textbf {\bibinfo {volume} {83}},\ \bibinfo {pages} {127502}
  (\bibinfo {year} {2011})},\ \Eprint {http://arxiv.org/abs/1102.2451}
  {arXiv:1102.2451 [gr-qc]} \BibitemShut {NoStop}%
\bibitem [{\citenamefont {Panosso~Macedo}(2024)}]{PanossoMacedo:2023qzp}%
  \BibitemOpen
  \bibfield  {author} {\bibinfo {author} {\bibfnamefont {R.}~\bibnamefont
  {Panosso~Macedo}},\ }\bibfield  {title} {\enquote {\bibinfo {title}
  {{Hyperboloidal approach for static spherically symmetric spacetimes: a
  didactical introductionand applications in black-hole physics}},}\ }\href
  {\doibase 10.1098/rsta.2023.0046} {\bibfield  {journal} {\bibinfo  {journal}
  {Phil. Trans. Roy. Soc. Lond. A}\ }\textbf {\bibinfo {volume} {382}},\
  \bibinfo {pages} {20230046} (\bibinfo {year} {2024})},\ \Eprint
  {http://arxiv.org/abs/2307.15735} {arXiv:2307.15735 [gr-qc]} \BibitemShut
  {NoStop}%
\bibitem [{\citenamefont {Panosso~Macedo}\ and\ \citenamefont
  {Zenginoglu}(2024)}]{PanossoMacedo:2024nkw}%
  \BibitemOpen
  \bibfield  {author} {\bibinfo {author} {\bibfnamefont {R.}~\bibnamefont
  {Panosso~Macedo}}\ and\ \bibinfo {author} {\bibfnamefont {A.}~\bibnamefont
  {Zenginoglu}},\ }\bibfield  {title} {\enquote {\bibinfo {title}
  {{Hyperboloidal approach to quasinormal modes}},}\ }\href {\doibase
  10.3389/fphy.2024.1497601} {\bibfield  {journal} {\bibinfo  {journal} {Front.
  in Phys.}\ }\textbf {\bibinfo {volume} {12}},\ \bibinfo {pages} {1497601}
  (\bibinfo {year} {2024})},\ \Eprint {http://arxiv.org/abs/2409.11478}
  {arXiv:2409.11478 [gr-qc]} \BibitemShut {NoStop}%
\bibitem [{\citenamefont {Ansorg}\ and\ \citenamefont
  {Panosso~Macedo}(2016)}]{Ansorg:2016ztf}%
  \BibitemOpen
  \bibfield  {author} {\bibinfo {author} {\bibfnamefont {M.}~\bibnamefont
  {Ansorg}}\ and\ \bibinfo {author} {\bibfnamefont {R.}~\bibnamefont
  {Panosso~Macedo}},\ }\bibfield  {title} {\enquote {\bibinfo {title}
  {{Spectral decomposition of black-hole perturbations on hyperboloidal
  slices}},}\ }\href {\doibase 10.1103/PhysRevD.93.124016} {\bibfield
  {journal} {\bibinfo  {journal} {Phys. Rev. D}\ }\textbf {\bibinfo {volume}
  {93}},\ \bibinfo {pages} {124016} (\bibinfo {year} {2016})},\ \Eprint
  {http://arxiv.org/abs/1604.02261} {arXiv:1604.02261 [gr-qc]} \BibitemShut
  {NoStop}%
\bibitem [{\citenamefont {Panosso~Macedo}\ and\ \citenamefont
  {Ansorg}(2014)}]{PanossoMacedo:2014dnr}%
  \BibitemOpen
  \bibfield  {author} {\bibinfo {author} {\bibfnamefont {R.}~\bibnamefont
  {Panosso~Macedo}}\ and\ \bibinfo {author} {\bibfnamefont {M.}~\bibnamefont
  {Ansorg}},\ }\bibfield  {title} {\enquote {\bibinfo {title} {{Axisymmetric
  fully spectral code for hyperbolic equations}},}\ }\href {\doibase
  10.1016/j.jcp.2014.07.040} {\bibfield  {journal} {\bibinfo  {journal} {J.
  Comput. Phys.}\ }\textbf {\bibinfo {volume} {276}},\ \bibinfo {pages} {357}
  (\bibinfo {year} {2014})},\ \Eprint {http://arxiv.org/abs/1402.7343}
  {arXiv:1402.7343 [physics.comp-ph]} \BibitemShut {NoStop}%
\bibitem [{\citenamefont {Bourg}\ \emph {et~al.}(2025)\citenamefont {Bourg},
  \citenamefont {Panosso~Macedo}, \citenamefont {Spiers}, \citenamefont
  {Leather}, \citenamefont {B{\'e}atrice},\ and\ \citenamefont
  {Pound}}]{Bourg:2025lpd}%
  \BibitemOpen
  \bibfield  {author} {\bibinfo {author} {\bibfnamefont {P.}~\bibnamefont
  {Bourg}}, \bibinfo {author} {\bibfnamefont {R.}~\bibnamefont
  {Panosso~Macedo}}, \bibinfo {author} {\bibfnamefont {A.}~\bibnamefont
  {Spiers}}, \bibinfo {author} {\bibfnamefont {B.}~\bibnamefont {Leather}},
  \bibinfo {author} {\bibfnamefont {B.}~\bibnamefont {B{\'e}atrice}}, \ and\
  \bibinfo {author} {\bibfnamefont {A.}~\bibnamefont {Pound}},\ }\bibfield
  {title} {\enquote {\bibinfo {title} {{Quadratic quasinormal modes at null
  infinity on a Schwarzschild spacetime}},}\ }\href {\doibase
  10.1103/fbz4-qsvn} {\bibfield  {journal} {\bibinfo  {journal} {Phys. Rev. D}\
  }\textbf {\bibinfo {volume} {112}},\ \bibinfo {pages} {044049} (\bibinfo
  {year} {2025})},\ \Eprint {http://arxiv.org/abs/2503.07432} {arXiv:2503.07432
  [gr-qc]} \BibitemShut {NoStop}%
\bibitem [{\citenamefont {Panosso~Macedo}\ \emph {et~al.}(2022)\citenamefont
  {Panosso~Macedo}, \citenamefont {Leather}, \citenamefont {Warburton},
  \citenamefont {Wardell},\ and\ \citenamefont
  {Zengino{\u{g}}lu}}]{PanossoMacedo:2022fdi}%
  \BibitemOpen
  \bibfield  {author} {\bibinfo {author} {\bibfnamefont {R.}~\bibnamefont
  {Panosso~Macedo}}, \bibinfo {author} {\bibfnamefont {B.}~\bibnamefont
  {Leather}}, \bibinfo {author} {\bibfnamefont {N.}~\bibnamefont {Warburton}},
  \bibinfo {author} {\bibfnamefont {B.}~\bibnamefont {Wardell}}, \ and\
  \bibinfo {author} {\bibfnamefont {A.}~\bibnamefont {Zengino{\u{g}}lu}},\
  }\bibfield  {title} {\enquote {\bibinfo {title} {{Hyperboloidal method for
  frequency-domain self-force calculations}},}\ }\href {\doibase
  10.1103/PhysRevD.105.104033} {\bibfield  {journal} {\bibinfo  {journal}
  {Phys. Rev. D}\ }\textbf {\bibinfo {volume} {105}},\ \bibinfo {pages}
  {104033} (\bibinfo {year} {2022})},\ \Eprint
  {http://arxiv.org/abs/2202.01794} {arXiv:2202.01794 [gr-qc]} \BibitemShut
  {NoStop}%
\bibitem [{\citenamefont {Zhou}\ and\ \citenamefont
  {Panosso~Macedo}(2025)}]{Zhou:2025xta}%
  \BibitemOpen
  \bibfield  {author} {\bibinfo {author} {\bibfnamefont {Y.}~\bibnamefont
  {Zhou}}\ and\ \bibinfo {author} {\bibfnamefont {R.}~\bibnamefont
  {Panosso~Macedo}},\ }\bibfield  {title} {\enquote {\bibinfo {title}
  {{Limiting geometry and spectral instability in Schwarzschild{\textendash}de
  Sitter spacetimes}},}\ }\href {\doibase 10.1103/n948-jnfh} {\bibfield
  {journal} {\bibinfo  {journal} {Phys. Rev. D}\ }\textbf {\bibinfo {volume}
  {112}},\ \bibinfo {pages} {084063} (\bibinfo {year} {2025})},\ \Eprint
  {http://arxiv.org/abs/2507.05370} {arXiv:2507.05370 [gr-qc]} \BibitemShut
  {NoStop}%
\bibitem [{\citenamefont {Berti}\ \emph {et~al.}(2022)\citenamefont {Berti},
  \citenamefont {Cardoso}, \citenamefont {Cheung}, \citenamefont {Di~Filippo},
  \citenamefont {Duque}, \citenamefont {Martens},\ and\ \citenamefont
  {Mukohyama}}]{Berti:2022xfj}%
  \BibitemOpen
  \bibfield  {author} {\bibinfo {author} {\bibfnamefont {E.}~\bibnamefont
  {Berti}}, \bibinfo {author} {\bibfnamefont {V.}~\bibnamefont {Cardoso}},
  \bibinfo {author} {\bibfnamefont {M.~H.-Y.}\ \bibnamefont {Cheung}}, \bibinfo
  {author} {\bibfnamefont {F.}~\bibnamefont {Di~Filippo}}, \bibinfo {author}
  {\bibfnamefont {F.}~\bibnamefont {Duque}}, \bibinfo {author} {\bibfnamefont
  {P.}~\bibnamefont {Martens}}, \ and\ \bibinfo {author} {\bibfnamefont
  {S.}~\bibnamefont {Mukohyama}},\ }\bibfield  {title} {\enquote {\bibinfo
  {title} {{Stability of the fundamental quasinormal mode in time-domain
  observations against small perturbations}},}\ }\href {\doibase
  10.1103/PhysRevD.106.084011} {\bibfield  {journal} {\bibinfo  {journal}
  {Phys. Rev. D}\ }\textbf {\bibinfo {volume} {106}},\ \bibinfo {pages}
  {084011} (\bibinfo {year} {2022})},\ \Eprint
  {http://arxiv.org/abs/2205.08547} {arXiv:2205.08547 [gr-qc]} \BibitemShut
  {NoStop}%
\bibitem [{\citenamefont {Kyutoku}\ \emph {et~al.}(2023)\citenamefont
  {Kyutoku}, \citenamefont {Motohashi},\ and\ \citenamefont
  {Tanaka}}]{Kyutoku:2022gbr}%
  \BibitemOpen
  \bibfield  {author} {\bibinfo {author} {\bibfnamefont {K.}~\bibnamefont
  {Kyutoku}}, \bibinfo {author} {\bibfnamefont {H.}~\bibnamefont {Motohashi}},
  \ and\ \bibinfo {author} {\bibfnamefont {T.}~\bibnamefont {Tanaka}},\
  }\bibfield  {title} {\enquote {\bibinfo {title} {{Quasinormal modes of
  Schwarzschild black holes on the real axis}},}\ }\href {\doibase
  10.1103/PhysRevD.107.044012} {\bibfield  {journal} {\bibinfo  {journal}
  {Phys. Rev. D}\ }\textbf {\bibinfo {volume} {107}},\ \bibinfo {pages}
  {044012} (\bibinfo {year} {2023})},\ \Eprint
  {http://arxiv.org/abs/2206.00671} {arXiv:2206.00671 [gr-qc]} \BibitemShut
  {NoStop}%
\bibitem [{Note2()}]{Note2}%
  \BibitemOpen
  \bibinfo {note} {While finishing this work, Ref.~\cite {Cao:2025afs} also
  comments on a similar result.}\BibitemShut {Stop}%
\bibitem [{\citenamefont {Gajic}\ and\ \citenamefont
  {Warnick}(2024)}]{Gajic:2024xrn}%
  \BibitemOpen
  \bibfield  {author} {\bibinfo {author} {\bibfnamefont {D.}~\bibnamefont
  {Gajic}}\ and\ \bibinfo {author} {\bibfnamefont {C.~M.}\ \bibnamefont
  {Warnick}},\ }\bibfield  {title} {\enquote {\bibinfo {title} {{Quasinormal
  modes on Kerr spacetimes}},}\ }\href@noop {} {\  (\bibinfo {year} {2024})},\
  \Eprint {http://arxiv.org/abs/2407.04098} {arXiv:2407.04098 [gr-qc]}
  \BibitemShut {NoStop}%
\bibitem [{\citenamefont {Rodrigo
  Panosso~Macedo}(2026)}]{rodrigo_panosso_macedo_2026_20177185}%
  \BibitemOpen
  \bibfield  {author} {\bibinfo {author} {\bibfnamefont {K.-i. K. H.~M.}\
  \bibnamefont {Rodrigo Panosso~Macedo}, \bibfnamefont {Takuya~Katagiri}},\
  }\href {\doibase 10.5281/zenodo.20177185} {\enquote {\bibinfo {title}
  {Hypebhpt/qnm\_resonances\_ep: Codes used in published version},}\ }
  (\bibinfo {year} {2026})\BibitemShut {NoStop}%
\end{thebibliography}%

\end{document}